\documentclass[twocolumn,aps,prl,superscriptaddress,showpacs]{revtex4-1}
\usepackage{graphicx}
\usepackage{color}
\usepackage{amsmath}
\usepackage{latexsym}
\usepackage{amsfonts}
\usepackage{amssymb}
\usepackage{array}
\usepackage{epsfig}
\begin{document}
\title{Maximum possible fidelity in $1\rightarrow 2$ qubits cloning}

\author{D. Gangopadhyay}
\affiliation{Department of Physics, Ramakrishna Mission Vivekananda University, Belur Matth, Howrah}
\author{A. Sinha Roy}
\affiliation{Department of Physics, Ramakrishna Mission Vivekananda University, Belur Matth, Howrah}
\date{\today}
\begin{abstract}
We re-analyse the Bu\v{z}ek-Hillery Universal Quantum Cloning machine protocol and show that it allows better values for fidelity and Hilbert-Schmidt norm than hitherto reported. This higher value for the fidelity is identical to the maximum fidelity of phase covariant quantum cloning of Bru\ss -Cinchetti-D'Ariano-Macchiavello. This  value of fidelity has also been obtained by  Niu and Griffiths in their work without machine states. This is the maximum possible fidelity obtainable in $1\rightarrow 2$ qubits cloning. We then describe a different and new state dependent cloning protocol with four machine states where all non-exact copies of input states are taken into account in the output and we use the Hessian method of determining extrema of multivariate functions. The fidelity for the best overall quantum cloning in this protocol is $\bar{F}=0.847$ with an associated von-Neumann entropy of $\bar{S}=0.825$.
\end{abstract}
\maketitle
\section*{Introduction}
Wootters, Zurek and Dieks \cite{wootters} first questioned whether it is possible to exactly clone a quantum state, i.e. produce copies of a quantum system each having the same state as the original. They came up with the famous {\it no cloning theorem} which states that it is impossible to perfectly clone an arbitrary quantum state $|\psi\rangle=\alpha|0\rangle +\beta |1\rangle$ where $|0\rangle$ and $|1\rangle$ are qubit states. Subsequently, Bu\v{z}ek and Hillery described a copying process which is input-state independent using a universal quantum copying machine \cite{buzek}. Gisin \cite{gisin} showed that the  Bu\v{z}ek-Hillery machine gives maximum fidelity for any arbitrary state $|\psi\rangle$. Later various quantum cloning machine protocols were introduced \cite{gis,scar}. Experimentally, quantum cloning machines have been implemented in quantum optics \cite{lamas} and  nuclear magnetic resonance systems \cite{du}.
 
In order to find the maximum fidelity (defined below in Section 2) obtainable in quantum cloning we must remember that the extremisation has to be with respect to the free parameters available. Maximum fidelity corresponds to minimum Hilbert Schmidt norm and \textit{vice versa}. In the  Bu\v{z}ek-Hillery scenario, the extremisation is input state independent, i.e. the Hilbert Schmidt norm is independent of $\alpha$ and $\beta$. The other parameters were the overlaps of the various machine states. Unitarity of the cloning transformation gave relations between the overlaps of the machine states. Using those relations the Cauchy-Schwarz inequality (CSI)bounds for the overlaps of machine states were determined. Imposing input state independence, Bu\v{z}ek and Hillery then obtained the values of these overlaps and showed that they satisfy the CSI bounds. Subsequently they used these results to obtain the best possible cloning.  

The motive of this work is to investigate whether the quantum cloning which was input state independent in the Bu\v{z}ek-Hillery protocol can be improved. We then compare our results with the standard works \cite{brub, niu}. We also consider a new state dependent quantum cloning protocol with four machine states where all non-exact copies of input states are taken into account in the output. We use the Hessian method and calculate fidelity, Hilbert-Schmidt norm and von-Neumann entropy of this new protocol. Our extremisation procedure is with respect to the overlaps of machine states. 

Accordingly, the plan of the paper is as follows. Below in the next section, we reconsider the Bu\v{z}ek and Hillery quantum cloning protocol and show that one can obtain a better value of the Hilbert-Schmidt norm and fidelity (defined below in Section 2) than hitherto known. In section 3 we reconsider Bru\ss, Cinchetti \textit{et al.} phase covariant quantum cloning protocol with input pure state as $|\psi\rangle_{a} =\alpha | 0\rangle_{a} +\beta | 1\rangle_{a}$ where $\alpha$ and $\beta$ are complex. In section 4 we consider  a state dependent quantum cloning protocol and obtain the relevant density operators and the relevant CSI  bounds for the overlaps of the machine states.  In section 5 we consider various possible choices for the overlaps of machine states. For certain choices of overlaps we get the best possible cloning in our protocol. The conclusions are in section 6.

\section{Bu\v{z}ek-Hillery UQCM protocol}
Bu\v{z}ek and Hillery proposed the Univeral Quantum Copying Machine (UQCM) protocol as \cite{buzek}
\begin{eqnarray}
\label{100}
| 0\rangle_{a}| Q\rangle_{x} \rightarrow
 | 0\rangle_{a}| 0\rangle_{b}| Q_0\rangle_{x}
+\Big[|0\rangle_{a}| 1\rangle_{b}+
| 1\rangle_{a}| 0\rangle_{b}
\Big]| Y_0\rangle_{x}\nonumber\\
\end{eqnarray}
 \begin{eqnarray}
 \label{101}
| 1\rangle_{a}| Q\rangle_{x} \rightarrow
| 1\rangle_{a}| 1\rangle_{b}| Q_1\rangle_{x}
+\Big[| 0\rangle_{a}| 1\rangle_{b}+
| 1\rangle_{a}| 0\rangle_{b}\Big]| Y_1\rangle_{x}.\nonumber\\
\end{eqnarray}
Here the subscript $a$ stands for states of original system and $b$ for
states of copied system. $| Q\rangle_{x}$ is input state of the copying machine, while $| Q_i\rangle_{x}$ and $| Y_i\rangle_{x}$ ( $i=0, 1$) are the final output states of the
copying machine. The relevant output density operator of $a$ mode is \cite{buzek}:
 \begin{eqnarray}
\hat{\rho}^{(out)}_{a} &= | 0\rangle_a\, _a\langle 0| \left[ \alpha^2 +\left(\beta^2 \,_x\langle Y_1|Y_1\rangle_x -\alpha^2 \,_x\langle Y_0|Y_0\rangle_x\right)\right]\nonumber\\
&+| 0\rangle_a\, _a\langle 1|\alpha\beta \left[
_x\langle Q_1|Y_0\rangle_x+\, _x\langle Y_1|Q_0\rangle_x \right]\nonumber\\
&+| 1\rangle_a\, _a\langle 0|\alpha\beta \left[\,
_x\langle Q_0|Y_1\rangle_x+\, _x\langle Y_0|Q_1\rangle_x \right]\nonumber\\
&+| 1\rangle_a\, _a\langle 1| \left[ \beta^2 +\alpha^2\,
_x\langle Y_0|Y_0\rangle_x -\beta^2\, _x\langle Y_1|Y_1\rangle_x \right]
\end{eqnarray}
Now the input density operator of the mode $a$ is 
\begin{eqnarray}
&&\hat{\rho}^{(id)}_{a} = \nonumber\\
&&\alpha^2 |0\rangle_{a}\,_{a}\langle 0|
+\alpha\beta |0\rangle_{a}\,_{a}\langle 1| +\beta\alpha |1\rangle_{a}\,_{a}\langle 0|+\beta^2 |1\rangle_{a}\,_{a}\langle 1|
\end{eqnarray}
Here the density operator of the output state of $a$ mode is different from the input state density operator of $a$ mode. This means during cloning (\textit{unitary transformation}) original input state is disturbed. To quantify the amount of disturbance the Hilbert-Schmidt norm of $a$ mode is defined as \cite{buzek}:
$D_a \equiv {\rm Tr}\left[\hat{\rho}^{(id)}_{a}-\hat{\rho}^{(out)}_{a}\right]^2$.
Another measure of distinguishability between two quantum states is fidelity \cite{sch}: $F={\rm Tr}\left(\sqrt{\hat{\rho}_{a}^{(id)}}\hat{\rho}_{a}^{(out)}\sqrt{\hat{\rho}_{a}^{(id)}}
\right)^{1/2}$.
Large $F$ means the states are less distinguishable.
For  Bu\v{z}ek-Hillery UQCM protocol, Hilbert Schmidt norm is \cite{buzek}
\begin{equation}
\label{cl30}
D_{a}=2A^{2}(4\alpha^{4}-4\alpha^{2}+1)+2\alpha^{2}(1-\alpha^{2})(1-2C)^{2}
\end{equation}
Here $\xi$ and $\eta/2$ in Bu\v{z}ek-Hillery paper are replaced by $A$ and $C$ in our calculation.
$_x\langle Y_0|Y_0\rangle_x=\,_x\langle Y_1|Y_1\rangle_x=A$ and $_x\langle Y_1|Q_0\rangle_x=\,_x\langle Y_0|Q_1\rangle_x=C$. The CSI bounds for $A$ and $C$ are $0\leq A\leq \frac{1}{2}$ and $0\leq C\leq1/2\sqrt{2}$ respectively. Both terms in the expression of $D_{a}$ are positive definite. $D_{a}$ is minimum with respect to the inner products of machine states $A$ and $C$ when $A=0$ and $C=1/2$ and the value of this minimum is $0$ . But $A$ can never be $0$ because then both $|Y_{0}\rangle$, $|Y_{1}\rangle$ will be zero. Also if $A=0$ then $C=0$ and things become meaningless. Further, $C=1/2$ violates CSI which give the bounds as $0\leq C\leq1/2\sqrt{2}$. So $C\neq1/2$. 
 
If we want $D_{a}$ to be input state independent, then $\frac{\partial}{\partial\alpha^{2}} D_{a}=0$ as in \cite{buzek}. From there we get  $A=\frac{1}{2}-C$. Then equation (\ref{cl30}) reduces to $D_{a}=\frac{(1-2C)^{2}}{2}$. So $D_{a}$ is minimum when $C$ is maximum. Here also $D_{a}=0$ if $C=1/2$. But this value of $C$ is ruled out for reasons already given in the previous paragraph. 

However, the Bu\v{z}ek-Hillery protocol can give better values for fidelity and Hilbert-Schmidt norm as we now show. If we choose the value of one of the overlaps within the CSI bound then the value of the other overlap is automatically fixed. Let us choose the maximum value of the overlap $C$ allowed by CSI which is $C=\frac{1}{2\sqrt{2}}$. Then $D_{a}$ would be minimum. For this value of $C$,  $A=\frac{1}{2}(1-\frac{1}{\sqrt{2}})$ and this value is within the CSI bound. For this set of values of the overlaps $D_{a}$ becomes minimum which is $D_{a}=\frac{3-2\sqrt{2}}{4}=0.0429$. This is the minimum possible value of the distance $D_{a}$ which is input state independent and the overlaps of machine states satisfy CSI. Bu\v{z}ek and Hillery got the minimum values of $D_{a}$ as $D_{a}=\frac{1}{18}=0.0556$ \cite{buzek}. Here our estimated minimum value of $D_{a}$ is lower than that of found by Bu\v{z}ek and Hillery.
 
Similarly the  fidelity is also  higher. The value of the fidelity is 
$F=\sqrt{1-A}=\sqrt{1/2+C} =0.9239$, for $C=\frac{1}{2\sqrt{2}}$. The value of the fidelity estimated by Bu\v{z}ek and Hillery, was $\sqrt{\frac{5}{6}}=0.9129$. So our estimated value is higher.

Bu\v{z}ek and Hillery \cite{buzek} also evaluated 
\begin{eqnarray}
&&D_{ab}^{(2)} \equiv {\rm Tr}\left[\hat{\rho}^{(id)}_{ab}-\hat{\rho}^{(out)}_{ab}\right]^2\nonumber\\
&&=1+8\alpha ^4\beta ^4 -4\alpha ^2 \beta ^2(1+2A)+(1-2A)^2 \nonumber\\
&&- 2(1-2A)(1-\alpha^ 2\beta ^2) +4 A^2
\end{eqnarray}
and found it's minimum value to be equal to $\frac{2}{9}=0.2222$.
Evaluating the same for $A=\frac{1}{2}(1-\frac{1}{\sqrt{2}})$ , (after averaging over $\alpha$) we get the minimum value of $\bar{D}_{ab}^{(2)}$ as$\frac{37}{15}-\frac{8\sqrt{2}}{5}=0.2039$. So we have a better estimate of $\bar{D}_{ab}^{(2)}$ also. Summarising 
\begin{table}[ht]
\caption{Comparison}
\centering
\begin{tabular}{c c c}
\hline\hline
Quantity & Bu\v{z}ek-Hillery value & Improved Bu\v{z}ek-Hillery value \\[0.5ex]
\hline
$D_{a}$ & 0.0556 & 0.0429 \\
Fidelity & 0.9129 & 0.9239 \\
$D_{ab}^{(2)}$ & 0.2222 & 0.2039 \\ [1ex]
\hline
\end{tabular}
\label{table:comparison}
\end{table}

Bu\v{z}ek and Hillery \cite{buzek} got the values of the various inner products of machine states as
$$
_x\langle Q_i|Q_i\rangle_x =2/3;\qquad
 _x\langle Y_i|Y_i\rangle_x =1/6; \qquad i=0,1
$$
$$
_x\langle Y_1|Y_0\rangle_x = \, _x\langle Q_1|Q_0\rangle_x =0;
$$
$$
_x\langle Y_0|Q_1\rangle_x = \, _x\langle Y_1|Q_0\rangle_x =1/3.
$$

For these values of inner products of machine states the cloning machine protocol (\ref{100}) and (\ref{101}) becomes \cite{buzek}
\begin{eqnarray}
\label{102}
| 0\rangle_{a}| Q\rangle_{x} \rightarrow
 \sqrt{\frac{2}{3}}| 0\rangle_{a}| 0\rangle_{b}|\uparrow \rangle_{x}
+\sqrt{\frac{1}{6}}\Big[|0\rangle_{a}| 1\rangle_{b}+
| 1\rangle_{a}| 0\rangle_{b}
\Big]|\downarrow \rangle_{x}\nonumber\\
\end{eqnarray}
 \begin{eqnarray}
 \label{103}
| 1\rangle_{a}| Q\rangle_{x} \rightarrow
\sqrt{\frac{2}{3}}| 1\rangle_{a}| 1\rangle_{b}|\uparrow \rangle_{x}
+\sqrt{\frac{1}{6}}\Big[| 0\rangle_{a}| 1\rangle_{b}+
| 1\rangle_{a}| 0\rangle_{b}\Big]|\downarrow \rangle_{x},\nonumber\\
\end{eqnarray}
where the initial machine state $|Q\rangle _{x}$ can be expressed as a linear superposition of two basis states $|\uparrow \rangle _{x}$ and $|\downarrow \rangle _{x}$.
In our calculation we get the values of various inner products of machine states as
$$
_x\langle Q_i|Q_i\rangle_x =\frac{1}{\sqrt{2}};\qquad
 _x\langle Y_i|Y_i\rangle_x =\frac{\sqrt{2}-1}{2\sqrt{2}}; \qquad i=0,1
$$
$$
_x\langle Y_1|Y_0\rangle_x = \, _x\langle Q_1|Q_0\rangle_x =0;
$$
$$
_x\langle Y_0|Q_1\rangle_x = \, _x\langle Y_1|Q_0\rangle_x =\frac{1}{2\sqrt{2}}.
$$

For these values of inner products of machine states the cloning machine protocol (\ref{100}) and (\ref{101}) becomes
\begin{eqnarray}
\label{104}
| 0\rangle_{a}| Q\rangle_{x} \longrightarrow
&&\sqrt{\frac{1}{\sqrt{2}}}| 0\rangle_{a}| 0\rangle_{b}|\uparrow \rangle_{x}\nonumber\\
&&+\sqrt{\frac{\sqrt{2}-1}{2\sqrt{2}}}\Big[|0\rangle_{a}| 1\rangle_{b}+
| 1\rangle_{a}| 0\rangle_{b}
\Big]|\downarrow \rangle_{x}
\end{eqnarray}
 \begin{eqnarray}
 \label{105}
| 1\rangle_{a}| Q\rangle_{x} \longrightarrow
&&\sqrt{\frac{1}{\sqrt{2}}}| 1\rangle_{a}| 1\rangle_{b}|\uparrow \rangle_{x}\nonumber\\
&&+\sqrt{\frac{\sqrt{2}-1}{2\sqrt{2}}}\Big[| 0\rangle_{a}| 1\rangle_{b}+
| 1\rangle_{a}| 0\rangle_{b}\Big]|\downarrow \rangle_{x}.
\end{eqnarray}

Now we consider the input pure state to be $|\psi\rangle_{a} =\alpha | 0\rangle_{a} +\beta | 1\rangle_{a}$ with $\alpha , \beta$ complex and $|\alpha |^{2}+|\beta |^{2}=1$. For this input state after cloning transformation (\ref{100}) and (\ref{101}) the output density operator of $a$ mode becomes
 \begin{eqnarray}
 \label{106}
\hat{\rho}^{(out)}_{a} = &| 0\rangle_a\, _a\langle 0| \left[ |\alpha |^2 +\left(|\beta |^2 \,_x\langle Y_1|Y_1\rangle_x -|\alpha |^2 \,_x\langle Y_0|Y_0\rangle_x\right)\right]\nonumber\\
&+| 0\rangle_a\, _a\langle 1|\alpha\beta ^{*} \left[
_x\langle Q_1|Y_0\rangle_x+\, _x\langle Y_1|Q_0\rangle_x \right]\nonumber\\
&+| 1\rangle_a\, _a\langle 0|\alpha ^{*}\beta \left[\,
_x\langle Q_0|Y_1\rangle_x+\, _x\langle Y_0|Q_1\rangle_x \right]+\nonumber\\
&| 1\rangle_a\, _a\langle 1| \left[ |\beta |^2 +\left(|\alpha |^2\,
_x\langle Y_0|Y_0\rangle_x-|\beta |^2\, _x\langle Y_1|Y_1\rangle_x\right)\right]\nonumber\\
\end{eqnarray}
where $*$ indicate complex conjugate. The fidelity in this case becomes
\begin{equation}
\label{107}
F=\sqrt{1-A+|\alpha |^{2}|\beta |^{2}(4C-2+4A)}
\end{equation}
 For fidelity to be input state independent $4C-2+4A=0$ \textit{i.e,} $A=\frac{1}{2}-C$ and the fidelity becomes $F=\sqrt{1-A}$. For the same argument as before we can easily see that the maximum value of fidelity is $F=\sqrt{1-A}=\sqrt{1/2+C} =0.9239$, for $C=\frac{1}{2\sqrt{2}}$.

So from above calculation it is proved that for any arbitrary input pure state $|\psi\rangle_{a} =\alpha | 0\rangle_{a} +\beta | 1\rangle_{a}$ with $\alpha , \beta$ real or complex the maximum value of fidelity is independent of input state with value $F=0.9239$ which is higher than that estimated by Bu\v{z}ek and Hillery. So, Bu\v{z}ek-Hillery Quantum Cloning Machine protocol is universal with higher value of fidelity than that estimated by Bu\v{z}ek and Hillery.

\section{Phase Covariant Cloning Machine}

Bru\ss, Cinchetti \textit{et al.} proposed a phase covariant quantum cloning protocol \cite{brub} as
\begin{eqnarray}
\label{108}
U|0\rangle _{a} |0\rangle _{b} |X\rangle _{x} =a|0\rangle _{a}|0\rangle _{b} |0\rangle _{x} +b(|0\rangle _{a}|1\rangle _{b}\nonumber\\
+|1\rangle {a}|0\rangle _{b})|1\rangle {x} +c|1\rangle _{a}|1\rangle _{b} |0\rangle _{x}\nonumber\\
\\
U|1\rangle _{a} |0\rangle _{b} |X\rangle _{x} =a|1\rangle _{a}|1\rangle _{b} |1\rangle _{x} +b(|0\rangle _{a}|1\rangle _{b}\nonumber\\
 +|1\rangle {a}|0\rangle _{b})|0\rangle {x} +c|1\rangle _{a}|1\rangle _{b} |1\rangle _{x} \nonumber
\end{eqnarray}

Subscript $a$ stands for states of original system, $b$ for states of copied system and $x$ for the machine states. Bru\ss, Cinchetti \textit{et al.} consider the input pure state as $|\psi\rangle_{a} =\alpha | 0\rangle_{a} +\beta | 1\rangle_{a}$ with $\alpha , \beta$ real and $\alpha ^{2}+\beta^{2}=1$. We consider most general situation \textit{i.e.} we take $\alpha$ and $\beta$ to be complex with $|\alpha |^{2}+|\beta |^{2}=1$. Here we consider $a$, $b$ and $c$ to be real. The unitarity of (\ref{108}) gives the relation
\begin{equation}
\label{109}
a^{2}+2b^{2}+c^{2}=1
\end{equation}

For the input state $|\psi\rangle_{a} =\alpha | 0\rangle_{a} +\beta | 1\rangle_{a}$ after cloning transformation (\ref{108}) the output density operator of $a$ mode becomes
\begin{eqnarray}
\label{110}
\hat{\rho} _{a}^{out}=&&|0\rangle\langle0|(a^{2}|\alpha |^{2}+b^{2}+c^{2}|\beta |^{2})+|0\rangle\langle 1|2(ab\alpha\beta ^{*}\nonumber\\
&&+bc\alpha ^{*}\beta)+|1\rangle\langle 0|2(ab\alpha ^{*}\beta+bc\alpha\beta ^{*})\nonumber\\
&&+|1\rangle\langle 1|(a^{2}|\beta |^{2}+b^{2}+c^{2}|\alpha |^{2})
\end{eqnarray}

Here $\hat{\rho} _{a}^{out}=\hat{\rho} _{b}^{out}$ \textit{i.e.} the states of the two modes $a$ and $b$ at the output of the copying machine are equal to each other. The fidelity of $a$ mode and $b$ mode are also equal to each other as
\begin{eqnarray}
\label{111}
F=\Big[a^{2}+b^{2}+2(2ab-a^{2}+c^{2})|\alpha |^{2}|\beta |^{2}\nonumber\\
+2bc(\alpha ^{2}{\beta ^{*}}^{2}+{\alpha ^{*}}^{2}\beta ^{2})\Big]^{1/2}
\end{eqnarray}
Now let us consider $\alpha = \alpha _{1}+i\alpha _{2}$, $\beta=\beta _{1}+i\beta _{2}$. Then fidelity becomes
\begin{eqnarray}
\label{112}
F=&&\Big[a^{2}+b^{2}+2(2ab+2bc-a^{2}+c^{2})(\alpha _{1}\beta _{1}+\alpha _{2}\beta _{2})^{2}\nonumber\\
&&+2(2ab-2bc-a^{2}+c^{2})(\alpha _{1}\beta _{2}-\alpha _{2}\beta _{1})^{2}\Big]^{1/2}
\end{eqnarray}

Now we consider 3 different possible situation.
\subsection{Case-1}
First of all we consider the input pure state such that $\alpha$ and $\beta$ are both real or both purely imaginary. In that case $(\alpha _{1}\beta _{1}+\alpha _{2}\beta _{2})^{2}$ is nonzero but $(\alpha _{1}\beta _{2}-\alpha _{2}\beta _{1})^{2}=0$. So fidelity becomes
\begin{equation}
\label{113}
F=\Big[a^{2}+b^{2}+2(2ab+2bc-a^{2}+c^{2})(\alpha _{1}\beta _{1}+\alpha _{2}\beta _{2})^{2}\Big]^{1/2}
\end{equation}
To make fidelity input state independent $2ab+2bc-a^{2}+c^{2}=0$ \textit{i.e.} $2b=a-c$. For $2b=a-c$ (\ref{109}) becomes
\begin{equation}
\label{114}
2a^{2}-4ab+6b^{2}=1
\end{equation}
Now if we maximize fidelity with the above constraint relation (\ref{108}) by use of Lagrange multipliers, the maximum fidelity becomes $F=\big[\frac{1}{2}+\sqrt{\frac{1}{8}}\big]^{1/2}=0.9239$ for $a=\frac{1}{2}+\sqrt{\frac{1}{8}}$, $b=\sqrt{\frac{1}{8}}$ and $c=\frac{1}{2}-\sqrt{\frac{1}{8}}$.

Incidentally Bru\ss, Cinchetti \textit{et al.} \cite{brub} have presented a constructive proof for the best $1\rightarrow 2$ cloning transformation acting on equatorial qubits. The question may be asked what is the status of the calculations in this work {\textit vis-a-vis} their work. Note that for the Bloch sphere a general state is written as $|\psi\rangle = cos~\frac{\theta}{2}|0\rangle + e^{i\phi}sin~\frac{\theta}{2}$. Here it should be noted that a general state  
$|\psi\rangle =\alpha | 0\rangle +\beta | 1\rangle$ with $\alpha , \beta$ complex basically involves $4$ 
real parameters. But the constraint $|\alpha|^{2} +|\beta|^{2} = 1$ means that only $3$ of these are independent. The Bloch sphere representation is a realisation of this where we have $3$ real parameters 
to consider as $\alpha = cos\frac{\theta}{2}$ is always real. So the results of \cite{brub} and the results of this section (i.e. $\alpha , \beta$ both real) are equivalent  (note that $\theta=\frac{\pi}{2}\equiv x-y\, plane$ and $\phi = 0 \equiv x-z\, plane$).    

\subsection{Case-2}
Here we consider the input pure state such that if $\alpha$ is real then $\beta$ is purely imaginary and vice versa . In that case $(\alpha _{1}\beta _{1}+\alpha _{2}\beta _{2})^{2}=0$ but $(\alpha _{1}\beta _{2}-\alpha _{2}\beta _{1})^{2}$ is nonzero. In this case fidelity becomes
\begin{equation}
F=\Big[a^{2}+b^{2}+2(2ab-2bc-a^{2}+c^{2})(\alpha _{1}\beta _{2}-\alpha _{2}\beta _{1})^{2}\Big]^{1/2}
\end{equation}

To make fidelity input state independent $2ab-2bc-a^{2}+c^{2}=0$ \textit{i.e.} $2b=a+c$. For $2b=a+c$ (\ref{109}) becomes (\ref{114}). If we maximize fidelity with the constraint relation (\ref{114}) by use of Lagrange multipliers, the maximum fidelity becomes $F=\big[\frac{1}{2}+\sqrt{\frac{1}{8}}\big]^{1/2}=0.9239$ for $a=\frac{1}{2}+\sqrt{\frac{1}{8}}$, $b=\sqrt{\frac{1}{8}}$ and $c=-\frac{1}{2}+\sqrt{\frac{1}{8}}$.

\subsection{Case-3}
In this case the input pure state is such that $\alpha$ and $\beta$ are both complex \textit{i.e.} they both have nonzero real and imaginary parts. Then $(\alpha _{1}\beta _{1}+\alpha _{2}\beta _{2})^{2}$ and $(\alpha _{1}\beta _{2}-\alpha _{2}\beta _{1})^{2}$ are both nonzero. The fidelity is given by (\ref{112}).

To make fidelity input state independent $2ab+2bc-a^{2}+c^{2}=0$ and $2ab-2bc-a^{2}+c^{2}=0$ \textit{i.e.} $2b=a-c$ and $2b=a+c$ both simultaneously satisfy. That means $c=0$ and $2b=a$. For $2b=a$ (\ref{109}) gives $b^{2}=\frac{1}{6}$ and the fidelity becomes $F=\sqrt{\frac{5}{6}}=0.9128$ for $a=\sqrt{\frac{2}{3}}$, $b=\sqrt{\frac{1}{6}}$.

From the above three possible choices of input state we see that for $\alpha$, $\beta$ to be either real or pure imaginary (\textit{i.e.} case (1) and (2)) we get maximum fidelity to be $F=\big[\frac{1}{2}+\sqrt{\frac{1}{8}}\big]^{1/2}=0.9239$. On the other hand for $\alpha$ and $\beta$ to be both complex (\textit{i.e.} case 3) the fidelity becomes $F=\sqrt{\frac{5}{6}}=0.9128$. so we can say that this cloning protocol is not universal because for any arbitrary pure state the cloning fidelity is not same.

\section{A new state dependent cloning protocol}
Recall the general quantum copying transformation rule for pure states on a two dimensional space \cite{buzek} :
\begin{eqnarray}
\label{1}
| 0\rangle_{a}| Q\rangle_{x} \longrightarrow \sum_{k,l=0}^{1}
| k\rangle_{a}| l\rangle_{b}| Q_{kl}\rangle_{x};
\end{eqnarray}
\begin{eqnarray}
| 1\rangle_{a}| Q\rangle_{x} \longrightarrow \sum_{m,n=0}^{1}
| m\rangle_{a}| n\rangle_{b}| Q_{mn}\rangle_{x},
\end{eqnarray}
where $| Q_{mn}\rangle_{x}$ are not necessarily orthonormal
for all possible values of $m$ and $n$. The general copying transformation
involves many free parameters
$_x\langle Q_{kl}| Q_{mn}\rangle_{x}$ characteristic of the copying machine.

We propose  the following protocol for the copying transformation:
\begin{eqnarray}
\label{cl1}
| 0\rangle_{a}| Q\rangle_{x} \longrightarrow
 &&| 0\rangle_{a}| 0\rangle_{b}| Q_0\rangle_{x}\nonumber\\
&&+\Big[|0\rangle_{a}| 1\rangle_{b}+
| 1\rangle_{a}| 0\rangle_{b}
+| 1\rangle_{a}| 1\rangle_{b}\Big]| Y_0\rangle_{x}
\end{eqnarray}
 \begin{eqnarray}
 \label{cl2}
| 1\rangle_{a}| Q\rangle_{x} \longrightarrow
&&| 1\rangle_{a}| 1\rangle_{b}| Q_1\rangle_{x}\nonumber\\
&&+\Big[| 0\rangle_{a}| 1\rangle_{b}+
| 1\rangle_{a}| 0\rangle_{b}+
| 0\rangle_{a}| 0\rangle_{b}\Big]| Y_1\rangle_{x}.
\end{eqnarray}
Here the subscript $a$ stands for states of original system and $b$ for
states of copied system. $| Q\rangle_{x}$ is input state of the copying machine, while $| Q_i\rangle_{x}$ and $| Y_i\rangle_{x}$ ( $i=0, 1$) are the final output states of the
copying machine. Our protocol choice is motivated by the fact that in a unitary copying 
transformation the kets $|0\rangle$ can transform into $|00\rangle,|01\rangle,|10\rangle$, and
$|11\rangle$ where $|00\rangle$ denotes the  perfectly copied state while the other remaining three states cannot qualify as perfectly copied states. Similarly,for the state $|1\rangle$ the perfectly copied outcome 
is $|11\rangle$ while $|00\rangle,|01\rangle,|10\rangle$ cannot be regarded as exact copies.
Since (\ref{cl1}), (\ref{cl2}) are unitary transformations,
\begin{eqnarray}
\label{cl3}
_x\langle Q_{i}| Q_{i}\rangle_{x} +
3\, _x\langle Y_{i}| Y_{i}\rangle_{x}=1;\qquad i=0,1
\end{eqnarray}
\begin{eqnarray}
\label{cl4}
_x\langle Y_{0}| Y_{1}\rangle_{x} =\,
_x\langle Y_{1}| Y_{0}\rangle_{x} =0.
\end{eqnarray}
Here we assume that the copying machine state vectors $|Q _i\rangle_{x}$ and $| Y_i\rangle_{x}$ are mutually orthogonal for simplicity:
\begin{eqnarray}
\label{cl5}
_x\langle Q_{i}| Y_{i}\rangle_{x} =0;\qquad i=0,1,
\end{eqnarray}
We see from equation (\ref{cl3}) that the machine states are not normalised to unity. There are also other overlaps of the machine states that will be important in our analysis. These are $_x\langle Y_0|Y_0\rangle_x=A$; $_x\langle Y_1|Y_1\rangle_x=B$; $_x\langle Y_1|Q_0\rangle_x=C$. Since the trace of the density operator is always unity, it will follow (as shown in the section 3) that $_x\langle Y_0|Q_1\rangle_x=-C$. Note that the coefficients $_x\langle Q_{0}| Q_{0}\rangle_{x}$ and $_x\langle Q_{1}| Q_{1}\rangle_{x}$ are related to $A$ and $B$ as $_x\langle Q_{0}| Q_{0}\rangle_{x}=1-3A$ and $_x\langle Q_{1}| Q_{1}\rangle_{x}=1-3B$.

Now consider an arbitrary input quantum state $| s\rangle_{a}$ :
\begin{eqnarray}
\label{cl6}
| s\rangle_{a} =\alpha | 0\rangle_{a} +\beta | 1\rangle_{a},
\end{eqnarray}
with $\alpha$, $\beta$ real and $\alpha^{2}+\beta^{2}=1$. Using (\ref{cl1}) and (\ref{cl2}), $| s\rangle_{a}$ becomes:
 \begin{eqnarray}
 \label{cl7}
|s\rangle_{a}|Q\rangle_{x}&&\longrightarrow \alpha|0\rangle_{a}|0\rangle_{b}|Q_{0}\rangle_{x}
+\alpha \Big[|0\rangle_{a}|1\rangle_{b}+|1\rangle_{a}|0\rangle_{b}\nonumber\\
&&+|1\rangle_{a}|1\rangle_{b}\Big]|Y_{0}\rangle_{x}+\beta|1\rangle_{a}|1\rangle_{b}|Q_{1}\rangle_{x}+\beta\Big[|0\rangle_{a}|1\rangle_{b}\nonumber\\
&&+|1\rangle_{a}|0\rangle_{b}+|0\rangle_{a}|0\rangle_{b}\Big]
|Y_{1}\rangle_{x}\equiv|\psi\rangle_{abx}^{(out)}
\end{eqnarray}
The density operator of the output mode $\hat{\rho}_{abx}^{(out)}\equiv |\Psi\rangle_{abx}^{(out)}\, _{abx}^{(out)}\langle\Psi |$ contains 16 terms (\ref{cl40}).
The reduced density operator of the $ab$-subsystem at the output  $\hat{\rho}^{(out)}_{ab} = {\rm Tr}_x \left[ \hat{\rho}^{(out)}_{abx}\right]$ also 
contains  16 terms (\ref{cl41}). The density operator of the original mode $a$ after copying can be obtained from the expression of $\hat{\rho}^{(out)}_{ab}$ by tracing over the mode $b$. Then the density operator of mode $a$ at output is (using expressions (\ref{cl3})-(\ref{cl5}))

\begin{eqnarray}
\label{cl8}
&&\hat{\rho}_{a}^{(out)}=|0\rangle_{a}\,_{a}\langle 0|[\alpha^{2}-2\alpha^{2}\,_{x}\langle Y_{0}|Y_{0}\rangle_{x} +2\beta^{2}\,_{x}\langle Y_{1}|Y_{1}\rangle_{x}\nonumber\\
&&+\alpha\beta\,_{x}\langle Y_{1}|Q_{0}\rangle_{x} +\alpha\beta\,_{x}\langle Q_{0}|Y_{1}\rangle_{x}]+|0\rangle_{a}\,_{a}\langle 1|[\alpha^{2}\,_{x}\langle Y_{0}|Y_{0}\rangle_{x}\nonumber\\
&&+\alpha\beta\,_{x}\langle Y_{0}|Q_{1}\rangle_{x} +\alpha\beta\,_{x}\langle Q_{0}|Y_{1}\rangle_{x}+\beta^{2}\,_{x}\langle Y_{1}|Y_{1}\rangle_{x}]\nonumber\\
&&+|1\rangle_{a}\,_{a}\langle 0|[\alpha^{2}\,_{x}\langle Y_{0}|Y_{0}\rangle_{x}
+\alpha\beta\,_{x}\langle Y_{0}|Q_{1}\rangle_{x} +\alpha\beta\,_{x}\langle Q_{0}|Y_{1}\rangle_{x}\nonumber\\
&&+\beta^{2}\,_{x}\langle Y_{1}|Y_{1}\rangle_{x}]+|1\rangle_{a}\,_{a}\langle 1|[2\alpha^{2}\,_{x}\langle Y_{0}|Y_{0}\rangle_{x} +\alpha\beta\,_{x}\langle Q_{1}|Y_{0}\rangle_{x}\nonumber\\
&&+\alpha\beta\,_{x}\langle Y_{0}|Q_{1}\rangle_{x} +\beta^{2}-2\beta^{2}\,_{x}\langle Y_{1}|Y_{1}\rangle_{x}],
\end{eqnarray}
Here all inner product terms are real, i.e. $_{x}\langle Y_{0}|Q_{1}\rangle_{x}=\,_{x}\langle Q_{1}|Y_{0}\rangle_{x}$, $_{x}\langle Y_{1}|Q_{0}\rangle_{x}=\,_{x}\langle Q_{0}|Y_{1}\rangle_{x}$. Then ${\rm Tr}\hat {\rho}_a= 1+2\alpha\beta[_{x}\langle Y_{0}|Q_{1}\rangle_{x}+\,_{x}\langle Y_{1}|Q_{0}\rangle_{x}]$. As the trace of a density operator is unity, so $_{x}\langle Y_{0}|Q_{1}\rangle_{x}=-\,_{x}\langle Y_{1}|Q_{0}\rangle_{x}$. Now using the notations for overlaps between the machine states introduced after equation (\ref{cl5}), \textit{viz.}, $A,B,C$, the last equation becomes:
\begin{eqnarray}
\label{cl10}
&&\hat{\rho}_{a}^{(out)}=|0\rangle_{a}\,_{a}\langle 0|[\alpha^{2} (1-2A)+2\alpha\beta C+2\beta^{2}B]\nonumber\\
&&+|0\rangle_{a}\,_{a}\langle 1|[\alpha^{2}A+\beta^{2}B]+|1\rangle_{a}\,_{a}\langle 0|[\alpha^{2}A+\beta^{2}B]\nonumber\\
&&+|1\rangle_{a}\,_{a}\langle 1|[\beta^{2}+2\alpha^{2}A-2\alpha\beta C-2\beta^{2}B]
\end{eqnarray}

If we express states $|0\rangle, |1\rangle$ by column vectors as $\begin{pmatrix}1\\0\end{pmatrix}$, and $\begin{pmatrix}0\\1\end{pmatrix}$ respectively, then the density operator of the original mode $a$ after copying becomes 
\begin{eqnarray}
\label{cl10}
&\hat{\rho}_{a}^{(out)}= \nonumber\\
&\begin{pmatrix}
 \alpha^{2}-2\alpha^{2}A+2\alpha\beta C+2\beta^{2}B & \alpha^{2}A+\beta^{2}B \\
\alpha^{2}A+\beta^{2}B & \beta^{2}+2\alpha^{2}A-2\alpha\beta C-2\beta^{2}B
\end{pmatrix}\nonumber\\
\end{eqnarray}
Note that from CSI we can find out bounds on the values of inner products A, B, C allowed by quantum mechanics. From CSI we can write
$_{x}\langle Q_{0}|Y_{0}\rangle_{x}^{2}\leq\,_{x}\langle Q_{0}|Q_{0}\rangle_{x} \,_{x}\langle Y_{0}|Y_{0}\rangle _{x}$. As we have shown before We re-analyse the Bu\v{z}ek-Hillery state independent Universal Quantum Cloning machine protocol and show that it allows 
better values for fidelity and Hilbert-Schmidt norm than hitherto reported. This higher value for the fidelity is 
identical to the maximum fidelity of phase covariant quantum cloning (i.e. state dependent cloning) of   
Bru\ss -Cinchetti-D'Ariano-Macchiavello. This value of fidelity has also been obtained by  Niu and Griffiths in their work 
without machine states.This is the maximum possible fidelity obtainable in $1\rightarrow 2$ qubit cloning. We then describe a different and new state dependent cloning protocol with four machine states where all non-exact copies of input states are taken into account in the output and we use the Hessian method of determining extrema of multivariate functions. The fidelity for the best overall quantum cloning in this protocol is $F=0.847$ with an associated von-Neumann entropy of $\bar{S}=0.825$.$_x\langle Q_{0}| Q_{0}\rangle_{x}=1-3A$, above inequality becomes $0\leq(1-3A)A$ which gives $0\leq A\leq\frac{1}{3}$. In similar way we can also find out the bounds of $B$ from the inequality of $_{x}\langle Q_{1}|Y_{1}\rangle_{x}^{2}\leq\,_{x}\langle Q_{1}|Q_{1}\rangle_{x}\,_{x}\langle Y_{1}|Y_{1}\rangle_{x}$, to be $0\leq B\leq \frac{1}{3}$. To find CSI bound for $C$ we consider that $_{x}\langle Y_{1}|Q_{0}\rangle_{x}^{2}\leq\,_{x}\langle Y_{1}|Y_{1}\rangle_{x}(1-3\,_{x}\langle Y_{0}|Y_{0}\rangle_{x}) \Longrightarrow C ^{2}\leq B(1-3A)$, from which we get the bound of C as $-\frac{1}{\sqrt{3}} \leq C\leq \frac{1}{\sqrt{3}}$.

Proceeding as before, we get $\hat{\rho}_{b}^{(out)}= \hat{\rho}_{a}^{(out)}$. This means that two density operators at the output are equal to each other. The input density operator 
of the mode $a$ is 
\begin{eqnarray}
\label{cl11}
\hat{\rho}^{(id)}_{a}= &&\alpha^2 |0\rangle_{a}\,_{a}\langle 0|
+\alpha\beta |0\rangle_{a}\,_{a}\langle 1| +\beta\alpha |1\rangle_{a}\,_{a}\langle 0|\nonumber\\
&&+\beta^2 |1\rangle_{a}\,_{a}\langle 1|.
\end{eqnarray}
which in matrix form becomes
\begin{equation}
\label{cl11}
\hat{\rho}_{a}^{(id)}=
\begin{pmatrix}
 \alpha^{2} & \alpha\beta \\
\alpha\beta & \beta^{2}
\end{pmatrix}
\end{equation}
However $\hat{\rho}_{a,b}^{(out)}\neq \hat{\rho}^{(id)}_{a}$.
This means that original input state is disturbed due to copying. 

\section{Hilbert-Schmidt norm, Fidelity and von Neumann entropy for various choices of overlaps of machine states}
%\label{sec:Hilbert-Schmidt norm Fidelity}
\subsection{A,B,C arbitrary}
The Hilbert-Schmidt norm corresponding to the operators (\ref{cl10}) and (\ref{cl11}) is
\begin{eqnarray}
\label{cl12}
D_{a}=&&2[5\alpha^{4}A^{2}+5\beta^{4}B^{2}+\alpha^{2}\beta^{2}\left( 1+4C^{2}-6AB\right)\nonumber\\
&&-2\alpha^{3}\beta A\left(1+4C\right)-2\alpha\beta^{3}B\left(1-4C\right)].
\end{eqnarray}
Note that using $\alpha^{2} + \beta^{2}=1$, one can rewrite the above equation in terms of $\alpha$ only.
We use the Hessian method of determining extrema of multivariate functions. Here we want to extremise with respect to the overlaps $A, B, C$. The Hessian matrix is  
\begin{eqnarray*}
\begin{pmatrix}
\frac{\partial^{2}D_{a}}{\partial A^{2}} & \frac{\partial^{2}D_{a}}{\partial A\partial B} & \frac{\partial^{2}D_{a}}{\partial A\partial C} \\
\frac{\partial^{2}D_{a}}{\partial B\partial A} & \frac{\partial^{2}D_{a}}{\partial B^{2}} & \frac{\partial^{2}D_{a}}{\partial B\partial C} \\
\frac{\partial^{2}D_{a}}{\partial C\partial A} & \frac{\partial^{2}D_{a}}{\partial C\partial B} & \frac{\partial^{2}D_{a}}{\partial C^{2}} 
\end{pmatrix}
=
\begin{pmatrix}
20\alpha^{4} & -12\alpha^{2}\beta^{2} & -16\alpha^{3}\beta \\
-12\alpha^{2}\beta^{2} & 20\beta^{4} & 16\alpha\beta^{3} \\
-16\alpha^{3}\beta & 16\alpha\beta^{3} & 16\alpha^{2}\beta^{2}
\end{pmatrix}
\end{eqnarray*}
The determinant of this matrix is $0$. So the minimum value of $D_{a}$ is indeterminate. This does not mean that there is no minimum. It only means that this cannot be determined in this scheme. From equation (\ref{cl12}) it is evident that our cloning procedure is input state dependent as $D_{a}$ depends on $\alpha$. If we do not specify which state to copy, i.e. the value of $\alpha$ is apriori unknown, then the next best option is averaging over $\alpha$ to get $\bar D_{a}$ and then minimizing this  with respect to 
overlaps of machine states.
\begin{eqnarray}
\label{cl13}
\bar{D_{a}}&&=\int\limits_0^1 D_{a}(\alpha)d\alpha \nonumber\\
&&=2\Big[A^{2}+\frac{8}{3}B^{2}+\frac{2}{15}\left( 1+4C^{2}-6AB\right)\nonumber\\
&&-\frac{4}{15}A\left(1+4C\right)-\frac{2}{5}B\left(1-4C\right)\Big].
\end{eqnarray}
We next minimize $\bar{D_{a}}$ using Hessian method as discussed above. The determinant of the Hessian matrix is positive. So $\bar{D_{a}}$ has a minimum and this is  $\bar{D}_{a}^{min}=\frac{157}{885}=0.177401$ for 
the values of $A$, $B$, $C$ as $A=\frac{13}{59}$, $B=\frac{9}{118}$, $C=\frac{25}{236}$.
The fidelity corresponding to the density operators (\ref{cl10}) and (\ref{cl11}) is
\begin{eqnarray}
\label{cl19}
F=&[(1-2A)\alpha^{4}+2(A+C)\alpha^{3}\beta +2(A+B)\alpha^{2}\beta^{2}\nonumber\\
&+2(B-C)\alpha\beta^{3}+(1-2B)\beta^{4}]^{1/2}.
\end{eqnarray}
Here we have use the formula for the square root of a two by two matrix 
$\begin{pmatrix}
A & B \\
C & D
\end{pmatrix}$
as
$\frac{\begin{pmatrix}
A+s & B \\
C & D+s
\end{pmatrix}}{t}
$, for $t\neq 0$, where $s=\pm \sqrt{d}$, $d$ is the determinant of the original matrix and $t=\pm \sqrt{T+2s}$ where $T=A+D$ be the trace of the original matrix.

Average value of the fidelity is
\begin{eqnarray}
\label{cl20}
\bar{F}=\int\limits_0^1 F(\alpha)d\alpha =\Big[\frac{11}{15}+\frac{2}{15}A-\frac{2}{5}B-\frac{2}{15}C\Big]^{1/2}.
\end{eqnarray}
The maximum  for $\bar{F}$ is $\bar{F} ^{max}=0.847$ for $A=13/59$, $B=9/118$ and $C=25/236$.

von-Neumann entropy for ideal input state is $S(\hat{\rho} _{a}^{(id)})=0$.
von-Neumann entropy for the density operator of the output state of $a$ mode, Eq.(\ref{cl10}) is \cite{chuang}
\begin{eqnarray}
\label{cl25}
S(\hat{\rho} _{a}^{(out)})=-\sum_{i=1}^{2}\lambda_{i}\ln\lambda_{i};\quad \lambda_{1,2}=\frac{1}{2}[1\pm K],
\end{eqnarray}
where
\begin{eqnarray}
\label{cl26}
K=&[1+8\alpha\beta C+(8B-4+16C^{2})\beta^{2}+16(2B-1)C\alpha\beta^{3}\nonumber\\
&+4(1-4B +5B^{2} -4C^{2})\beta^{4}-20\alpha^{2}A^{2}-8\alpha^{2}A(1\nonumber\\
&+4C\alpha\beta +(3B-2)\beta^{2})]^{1/2}.
\end{eqnarray}
In Eq.(\ref{cl25}) and Eq.(\ref{cl26}) substituting the values of $A=13/59$, $B=9/118$ and $C=25/236$ and taking the average value over $\alpha$, we get average value of von-Neumann entropy $\bar{S}=\int\limits_0^1 S(\alpha)d\alpha=0.825$. So, von-Neumann entropy of initial input state is zero and after copy, the average value of von-Neumann entropy is $0.8250$. As the von-Neumann entropy is not zero, So, during the copy, initial input pure state disturbed and becomes mixed state.
\subsection{A=B, C arbitrary}
Here $_x\langle Y_0|Y_0\rangle_x= _x\langle Y_1|Y_1\rangle_x$.
Then from Eq.(\ref{cl12})
\begin{eqnarray}
\label{cl14}
D_{a}=&[5A^{2}+\alpha^{2}\beta^{2}\left( 1+4C^{2}-16A^{2}\right)\nonumber\\
&-2\alpha\beta A\{1+4C\left(\alpha^{2}-\beta^{2}\right) \}]
\end{eqnarray}
Minimizing $D_{a}$ with respect to $A$ and $C$ in same way as before, we get
$D_{a}^{min}=0$ for $A=\alpha\beta$, $C=\alpha^{2}-\beta^{2}$. 
So $\hat{\rho}^{(out)}_{a}=\hat{\rho}^{(id)}_{a}$, i.e,  perfect cloning. Here 
$0\leq A\leq 0.5$ while $-1\leq C\leq 1$. 
But these bounds are different from those obtained using the CSI which were $0\leq A\leq\frac{1}{3}$, $-\frac{1}{{\sqrt 3}}\leq C\leq\frac{1}{{\sqrt 3}}$. Within this region there are no values of $\alpha$ and/or $\beta$, where $A$ and $C$ both satisfy the values obtained using  CSI. This is shown in figure (1) which clearly depicts the impossibility of perfect cloning of any arbitrary quantum state.

The fidelity in this case is using Eq.(\ref{cl19}) (for $A=B$)
\begin{eqnarray}
\label{cl21}
F=&&[1-2\alpha^{2}+2\alpha^{4}-2C\alpha\beta +4C\alpha^{3}\beta \nonumber\\
&&2A(4\alpha^{2}-1+\alpha\beta)]^{1/2}
\end{eqnarray}
Now for $A=\alpha\beta$ and $C=\alpha^{2}-\beta^{2}$, the fidelity becomes 1.
These values of $A,C$ are outside the CSI allowed interval as already seen.

For the situation $A=B$ Eq.(\ref{cl25}) becomes more simplified, and for 
$A=\alpha\beta$ and $C=\alpha^2-\beta^2$, the von-Neumann entropy vanishes.

Usually CSI violations in other branches of physics signal the onset of quantum behaviour. This is so when  the CSI is analysed in the context of operators in various situations connected with optics \cite{marino, walls}. 
But in our copying protocol, the CSIs are in the context of overlaps of machine states. 
Hence the implications are totally  different. CSI violations here imply that the machine states no longer 
belong to standard quantum mechanical Hilbert spaces which are metric spaces. CSI violations imply violation 
of the triangle inequality which is a basic property satisfied by all vectors in a metric space. Therefore,
results obtained by violating CSI ("perfect cloning") are unacceptable as quantum mechanics is being violated. 
Therefore, the {\it no cloning theorem} is further strengthened.
\begin{figure}
\resizebox{8.0cm}{2.5cm}{\includegraphics{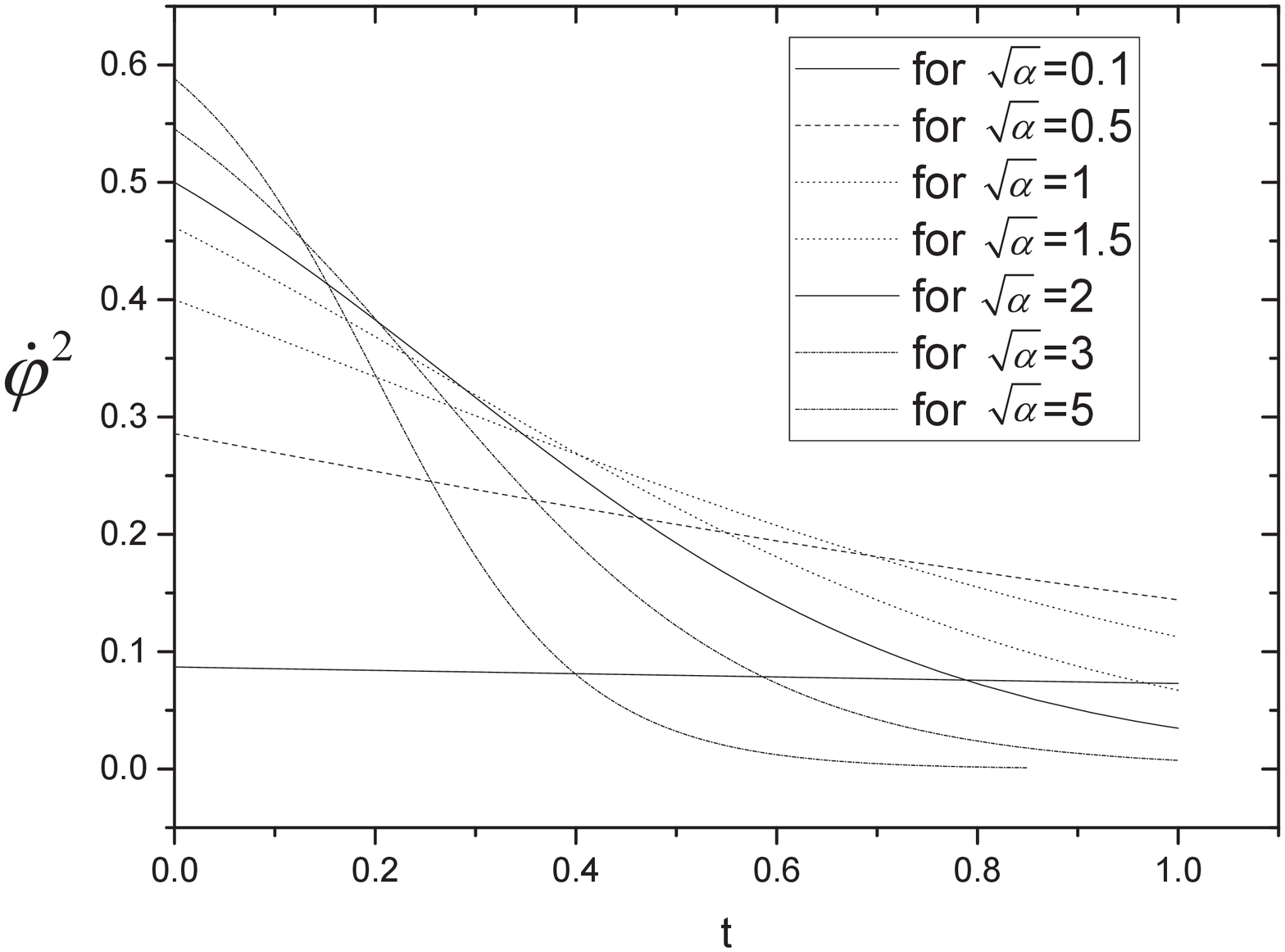}}
\caption{In Fig.1 we plot $A=\alpha\sqrt{1-\alpha ^{2}}$, $C=2\alpha^{2} -1$ for $\alpha$ from 0 to 1 which satisfy the range of A and C we got from Cauchy-Schwarz inequality. In the plot we see that for $\alpha$ from $0$ to $0.3568$ we get the plot of $A$ only. For $\alpha$ from $0.4597$ to $0.8881$ we get the plot of $C$ only and for $\alpha$ from $0.9342$ to $1$ we get the plot of $A$ only. So in the whole range of $\alpha$ there is no single value of $\alpha$ where we get the value of A and C together.}
\end{figure}

Since $D_{a}^{min}=0$ is unphysical,  we now determine $\bar {D_{a}}$.
\begin{eqnarray}
\label{cl15}
\bar{D_{a}}=2\Big[\frac{43}{15}A^{2}+\frac{2}{15}(1+4C^{2})-\frac{2}{3}A+\frac{8}{15}AC\Big]
\end{eqnarray}
Proceeding as before, $\bar{D_{a}}^{min}=\frac{38}{205}=0.185$ for $A=5/41$, $C=-5/82$.

Average value of fidelity $F$ over all possible values of $\alpha$: 
\begin{eqnarray}
\label{cl22}
\bar{F}=\Big[\frac{11}{15}-\frac{2}{15}C-\frac{4}{15}A\Big]^{1/2}
\end{eqnarray}
Then for $A=5/41$ and $C=-5/82$, $F^{max}=0.842$.

In this situation, i.e. for $A=5/41$ and $C=-5/82$, if we average over $\alpha$ 
one gets the average value of von-Neumann entropy of the output state of $a$ mode as 
$\bar{S}=0.8438$.

\subsection{A$\neq$B, C=0}
Here implication is $_x\langle Y_0|Y_0\rangle_x\neq\,_x\langle Y_1|Y_1\rangle_x$ and $_x\langle Y_1|Q_0\rangle_x=0$. From equation Eq.(\ref{cl10})
\begin{eqnarray}
\label{cl16}
\hat{\rho}^{(out)}_{a}=&&|0\rangle_{a}\,_{a}\langle 0|\big[\alpha^{2}-2\alpha^{2}A+2\beta^{2}B\big]+|0\rangle_{a}\,_{a}\langle 1|\big[\alpha^{2}A\nonumber\\
&&+\beta^{2}B\big]+|1\rangle_{a}\,_{a}\langle 0|\big[\alpha^{2}A+\beta^{2}B\big]\nonumber\\
&&+|1\rangle_{a}\,_{a}\langle 1|\big[\beta^{2}+2\alpha^{2}A-2\beta^{2}B\big].
\end{eqnarray}
The Hilbert-Schmidt norm corresponding to the density operators (\ref{cl11}) and (\ref{cl16}) is
\begin{eqnarray}
\label{cl17}
D_{a}=&&2\big[5\alpha^{4}A^{2}+5\beta^{4}B^{2}+\alpha^{2}\beta^{2}(1-6AB)\nonumber\\
&&-2\alpha\beta(\alpha^{2}A+\beta^{2}B)\big]
\end{eqnarray}
Here $D_{a}^{min}=0$ for $A=\alpha/2\beta$, $B=\beta/2\alpha$, i.e. perfect cloning. 
Since $0\leq\alpha,\beta\leq 1$ one can have $0\leq A,B\leq\infty$. But CSI's give
$0\leq A,B\leq \frac{1}{3}$. Hence CSI is violated. $A$ and $B$ cannot simultaneously satisfy the acceptable range of values of $\alpha$ and/or $\beta$ as shown in the figure (2). Therefore, for reasons already given before, we reject this result of perfect cloning.

\begin{figure}
\resizebox{8.0cm}{2.5cm}{\includegraphics{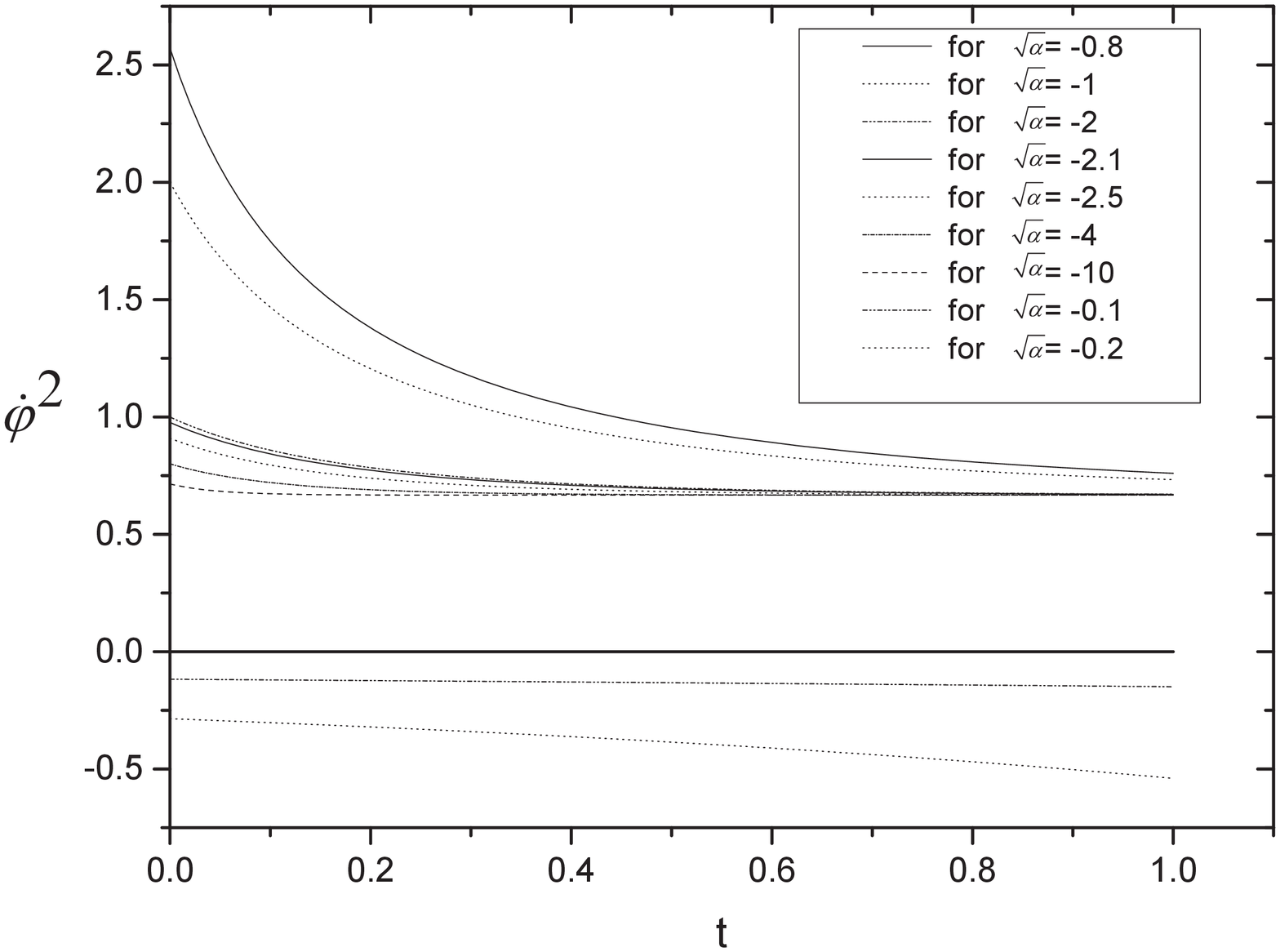}}
\caption{In fig.2 we plot $A=\alpha /2\sqrt{1-\alpha ^{2}}$, $B=2\sqrt{1-\alpha ^{2}}/\alpha$ for $\alpha$ from 0 to 1 which satisfy the range of A and B we got from CSI. In this figure for $\alpha$ from $0$ to $0.6546$ we get the plot of A only while for $\alpha$ from $0.8944$ to $1$ we get the plot of B only. So in the whole range of $\alpha$ there is no single value of $\alpha$ where we get the value of A and B together }
\end{figure}

For $A\neq B$ and $C=0$, Eq.(\ref{cl19}) becomes
\begin{eqnarray}
\label{cl23}
F=&[(1-2A)\alpha^{4}+2A\alpha^{3}\beta +2(A+B)\alpha^{2}\beta^{2}\nonumber\\
&+2B\alpha\beta^{3}+(1-2B)\beta^{4}]^{1/2}
\end{eqnarray}
For $A=\alpha/2\beta$ and $B=\beta/2\alpha$, the fidelity becomes 1. But from CSI 
we get different bounds for  $A$ and $B$ ($0\leq A\leq \frac{1}{3}$, $0\leq B\leq \frac{1}{3}$).

For $A=\alpha/2\beta$ and $B=\beta/2\alpha$, the von-Neumann entropy obtained from
 Eq.(\ref{cl25}) is zero.

So consider $\bar{D_{a}}$:
\begin{eqnarray}
\label{cl18}
\bar{D_{a}}=2\Big[A^{2}+\frac{8}{3}B^{2}+\frac{2}{15}(1-6AB)-\frac{4}{15}A-\frac{2}{5}B\Big]
\end{eqnarray}
We get $\bar{D_{a}}^{min}=0.1799$ for $A=\frac{49}{282}$, $B=\frac{19}{188}$. 

Average value of fidelity $F$ over all possible values of $\alpha$ from $0$ to $1$ is
\begin{eqnarray}
\label{cl24}
\bar{F}=&\Big[\frac{1}{5}(1-2A)+\frac{4}{15}(2A+B)+\frac{2}{5}B+\frac{8}{15}(1-2B)\Big]^{1/2}
\end{eqnarray}
and $\bar{F}^{max}=0.8462$ for $A=49/282$ and $B=19/188$.

For $A=49/282$, $B=19/188$, after averaging over $\alpha$ one has average value of von-Neumann entropy $\bar{S}=0.8297$.
\section{Conclusion}
%\label{sec:Conc}
\noindent
We have re-investigated the Bu\v{z}ek-Hillery Universal Quantum Cloning Machine and showed that it is possible to get a higher value of fidelity than that estimated by Bu\v{z}ek and Hillery. Maximum fidelity is obtained by choosing the maximum value of $C$ allowed by its CSI bounds. Input state independence of copying protocol implies that $A=\frac{1}{2}-C$. So the allowed maximum value of $C$ will correspond to the allowed minimum value of $A$ and this is also within the allowed CSI bounds for $A$. We have followed a slightly different route which is explained in section 2.

The  principal results of this work may be summarised as 

(1) We have  shown that for any arbitrary input pure state $|\psi\rangle_{a} =\alpha | 0\rangle_{a} +\beta | 1\rangle_{a}$ with $\alpha , \beta$ real or complex, the maximum value of fidelity in Bu\v{z}ek-Hillery quantum cloning machine is independent of input state i.e. Bu\v{z}ek-Hillery Quantum Cloning Machine protocol is universal.

(2)Our calculated maximum fidelity $F=0.9239$ is higher than that originally estimated by Bu\v{z}ek and Hillery  which was $F=0.9129$. Our Hilbert-Schmidt norm value is $D_{a}=0.0429$ whereas 
Bu\v{z}ek and Hillery obtained $D_{a}= 0.0556$.
 
(3)We have also re-investigated the Bru\ss, Cinchetti \textit{et al.} phase covariant quantum cloning protocol \cite{brub} with input pure state as $|\psi\rangle_{a} =\alpha | 0\rangle_{a} +\beta | 1\rangle_{a}$ where $\alpha$ and $\beta$ are complex. We find that for $\alpha$, $\beta$ to be either real or pure imaginary (\textit{i.e.} case (1) and (2) of section 3) we get maximum fidelity to be $F=\big[\frac{1}{2}+\sqrt{\frac{1}{8}}\big]^{1/2}=0.9239$. On the other hand for $\alpha$ and $\beta$  both complex with nonzero real and imaginary parts (\textit{i.e.} case (3) of section 3) the fidelity becomes $F=\sqrt{\frac{5}{6}}=0.9128$.

(4)It is interesting to note that the maximum possible value for the fidelity is identical in Bu\v{z}ek-Hillery universal quantum cloning machine and the phase covariant quantum cloning machine of Bru\ss -Cinchetti-D'Ariano-Macchiavello. Though this maximum fidelity can be achieved for any arbitrary input pure state in Bu\v{z}ek-Hillery cloning protocol but for Bru\ss -Cinchetti-D'Ariano-Macchiavello cloning protocol this is only possible for some input pure state (\textit{i.e.} case (1) and (2) of section 3)
and For other states (\textit{i.e.} case (3) of section 3) the maximum value of fidelity is lower. Therefore Bru\ss, Cinchetti \textit{et al.} cloning machine protocol is not universal. This maximum value of fidelity $F=0.9239$ has also been obtained by  Niu and Griffiths \cite{niu} in their work without machine states. Therefore, we can conclude that {\it upper bound of maximum possible fidelity in $1\rightarrow 2$ qubits cloning is $\sqrt{\frac{1}{2}(1+\frac{1}{\sqrt{2}})}=0.9239$.}

 (5)We have also described a new  input state dependent cloning protocol with four machine states where all non-exact copies of input states  are taken into account in the output. We have used the Hessian method of extremisation of multivariate functions. The extremisation procedure is with respect to the overlaps of machine states. In our copying protocol, we have investigated 
all possible choices for machine states and determined the values of inner products of machine states that give the best optimal cloning. The best overall quantum cloning is obtained in the first of the three choices. 
The Hilbert-Schmidt norm is $\bar{D_{a}}=0.1774$, the fidelity is $\bar{F}=0.847$ and the von-Neuman entropy is $\bar{S}=0.8250$. These values correspond to $A=13/59$, $B=9/118$ and $C=25/236$, where $A,B,C$ are the relevant overlaps of machine states.
\section{Acknowledgements}
\noindent
A. Sinha Roy thanks UGC-CSIR for providing a Research Fellowship Sr.No. 2061151173 under which this work was done.
\begin{widetext}
\section{appendix}
The density operator of the output mode,
\begin{eqnarray}
\label{cl40}
\hat{\rho}_{abx}^{(out)}
&&\equiv |\Psi\rangle_{abx}^{(out)}\, _{abx}^{(out)}\langle\Psi |\nonumber\\
&&=\alpha ^{2}|00\rangle \langle 00|(|Q_0\rangle_{x}\langle Q_0|)+\alpha ^{2}[|00\rangle \langle 01|+|00\rangle \langle 10|+|00\rangle \langle 00|](|Q_0\rangle_{x}\langle Y_0|)+\alpha\beta |00\rangle \langle 11|\nonumber\\
&&(|Q_0\rangle_{x}\langle Q_0|)+\alpha\beta [|00\rangle \langle 01|+|00\rangle \langle 10|+|00\rangle \langle 11|(|Q_0\rangle_{x}\langle Y_1|)+\alpha ^{2}[|01\rangle \langle 00|+|10\rangle \langle 00|\nonumber\\
&&+|11\rangle \langle 00|](|Y_0\rangle_{x}\langle Q_0|)+\alpha^{2}[(|01\rangle+|10\rangle+|11\rangle)(\langle 01|+\langle 10|+\langle 11|)](|Y_0\rangle_{x}\langle Y_0|)\nonumber\\
&&+\alpha\beta [|01\rangle \langle 11|+|10\rangle \langle 11|+|11\rangle \langle 11|](|Y_0\rangle_{x}\langle Q_1|)+\alpha\beta [(|01\rangle+|10\rangle+|11\rangle)\nonumber\\
&&(\langle 01|+\langle 10|+\langle 00|)](|Y_0\rangle_{x}\langle Y_1|)+\alpha\beta |11\rangle \langle 00|(|Q_0\rangle_{x}\langle Q_0|)+\alpha\beta[|11\rangle \langle 01|+|11\rangle \langle 10|\nonumber\\
&&+|11\rangle \langle 11|](|Q_1\rangle_{x}\langle Y_0|)+\beta^{2} |11\rangle \langle 11|(|Q_1\rangle_{x}\langle Q_1|)+\beta^{2}[|11\rangle \langle 01|+|11\rangle \langle 10|+|11\rangle \langle 00|]\nonumber\\
&&(|Q_1\rangle_{x}\langle Y_1|)+\alpha\beta [|01\rangle \langle 00|+|10\rangle \langle 00|+|00\rangle \langle 00|](|Y_1\rangle_{x}\langle Q_0|)+\alpha\beta [(|01\rangle+|10\rangle+|00\rangle)\nonumber\\
&&(\langle 01|+\langle 10|+\langle 11|)](|Y_1\rangle_{x}\langle Y_0|)+\beta^{2} [|01\rangle \langle 11|+|10\rangle \langle 11|+|00\rangle \langle 11|](|Y_1\rangle_{x}\langle Q_1|)\nonumber\\
&&\beta^{2}[(|01\rangle+|10\rangle+|00\rangle)(\langle 01|+\langle 10|+\langle 00|)](|Y_1\rangle_{x}\langle Y_1|)
\end{eqnarray}
 Where $|00\rangle\equiv |0\rangle _{a}|0\rangle _{b}$, $|01\rangle\equiv |0\rangle _{a}|1\rangle _{b}$, $|10\rangle\equiv |1\rangle _{a}|0\rangle _{b}$ and $|11\rangle\equiv |1\rangle _{a}|1\rangle _{b}$.The reduced density operator of the original-copy subsystem after copying procedure is
\begin{eqnarray}
\label{cl41}
\hat{\rho}^{(out)}_{ab} &&= {\rm Tr}_x \left[ \hat{\rho}^{(out)}_{abx}\right]\nonumber\\
&&=\alpha ^{2}|00\rangle \langle 00|(\langle Q_0|Q_0\rangle_{x})+\alpha ^{2}[|00\rangle \langle 01|+|00\rangle \langle 10|+|00\rangle \langle 00|](\langle Y_0|Q_0\rangle_{x})+\alpha\beta |00\rangle \langle 11|\nonumber\\
&&(\langle Q_1|Q_0\rangle_{x})+\alpha\beta [|00\rangle \langle 01|+|00\rangle \langle 10|+|00\rangle \langle 11|(\langle Y_1|Q_0\rangle_{x})+\alpha ^{2}[|01\rangle \langle 00|+|10\rangle \langle 00|\nonumber\\
&&+|11\rangle \langle 00|](\langle Q_0|Y_0\rangle_{x})+\alpha^{2}[(|01\rangle+|10\rangle+|11\rangle)(\langle 01|+\langle 10|+\langle 11|)](\langle Y_0|Y_0\rangle_{x})\nonumber\\
&&+\alpha\beta [|01\rangle \langle 11|+|10\rangle \langle 11|+|11\rangle \langle 11|](\langle Q_1|Y_0\rangle_{x})+\alpha\beta [(|01\rangle+|10\rangle+|11\rangle)\nonumber\\
&&(\langle 01|+\langle 10|+\langle 00|)](\langle y_1|Y_0\rangle_{x})+\alpha\beta |11\rangle \langle 00|(\langle Q_0|Q_1\rangle_{x})+\alpha\beta[|11\rangle \langle 01|+|11\rangle \langle 10|\nonumber\\
&&+|11\rangle \langle 11|](\langle Y_0|Q_1\rangle_{x})+\beta^{2} |11\rangle \langle 11|(\langle Q_1|Q_1\rangle_{x})+\beta^{2}[|11\rangle \langle 01|+|11\rangle \langle 10|+|11\rangle \langle 00|]\nonumber\\
&&(\langle Y_1|Q_1\rangle_{x})+\alpha\beta [|01\rangle \langle 00|+|10\rangle \langle 00|+|00\rangle \langle 00|](\langle Q_0|Y_1\rangle_{x})+\alpha\beta [(|01\rangle+|10\rangle+|00\rangle)\nonumber\\
&&(\langle 01|+\langle 10|+\langle 11|)](\langle Y_0|Y_1\rangle_{x})+\beta^{2} [|01\rangle \langle 11|+|10\rangle \langle 11|+|00\rangle \langle 11|](\langle Q_1|y_1\rangle_{x})\nonumber\\
&&\beta^{2}[(|01\rangle+|10\rangle+|00\rangle)(\langle 01|+\langle 10|+\langle 00|)](\langle Y_1|Y_1\rangle_{x}).
\end{eqnarray}
\end{widetext}

\end{document}